\documentclass{PoS}
\usepackage{bm}

\title{Improved automated lattice perturbation theory in background field gauge}

\ShortTitle{Improved automated lattice perturbation theory}

\author{T.C. Hammant, R.R. Horgan, C.J. Monahan\\
       DAMTP, CMS, University of Cambridge, Cambridge CB3 0WA, U.K.
}
\author{A.G. Hart\thanks{previous address: SUPA, School of Physics, University of Edinburgh, Edinburgh EH9 3JZ, U.K.}\\
       Cray Exascale Research Initiative, JCMB, King's Buildings, Edinburgh EH9 3JZ, U.K.
}
\author{E.H. M\"uller\thanks{new address: Meteorological Office, FitzRoy Road, Exeter, Devon EX1 3PB, U.K.}\\
       SUPA, School of Physics, University of Edinburgh, Edinburgh EH9 3JZ, U.K.}
\author{A. Gray, K. Sivalingham\\
       EPCC, JCMB, King's Buildings, Edinburgh EH9 3JZ, U.K.
}
\author{\speaker{G.M. von Hippel} \\
        Institut f\"ur Kernphysik, Johannes-Gutenberg-Universit\"at Mainz, 55099 Mainz, Germany\\
        \email{hippel@kph.uni-mainz.de}
        }

\abstract{We present an algorithm to automatically derive Feynman rules for lattice perturbation theory in background field gauge. Vertices with an arbitrary number of both background and quantum legs can be derived automatically from both gluonic and fermionic actions. The algorithm is a generalisation of our earlier algorithm based on prior work by L\"uscher and Weisz. We also present techniques allowing for the parallelisation of the evaluation of the often rather complex lattice Feynman rules that should allow for efficient implementation on GPUs, but also give a significant speed-up when calculating the derivatives of Feynman diagrams with respect to external momenta.}

\FullConference{The XXVIII International Symposium on Lattice Field Theory, Lattice2010\\
		June 14-19, 2010\\
		Villasimius, Italy}

\begin{document}

\newcommand{\rme}{\mathrm{e}}

\section{Automated Lattice Perturbation Theory}

While the lattice offers a fully nonperturbative regulator for Quantum Field
Theory, perturbative calculations still play an important role in
renormalising, matching and improving operators,
both those appearing in the action and those used to measure
observables. For the highly improved actions now widely used, such as the
Highly Improved Staggered Quark (HISQ) action 
\cite{HISQ}
or moving NRQCD,
\cite{mNRQCD}
manual derivation and implementation of the Feynman rules would be highly
impractical. Automated methods are therefore required. A general algorithm
to derive the Feynman rules for arbitrary traced closed Wilson loops was
derived by L\"uscher and Weisz in
\cite{LW}.
We have generalised this algorithm to include fermionic fields
\cite{Hart:2004}
and have implemented it in the HiPPy/HPsrc packages
\cite{Hart:2009}.

\subsection{The HiPPy package}

HiPPy is an is an automated tool for generating Feynman rules from arbitrary lattice actions, which is written entirely in {Python}
\cite{www_python}.
Starting from the perturbative expansion $U_{\mu}(\bm{x})=\mathrm{e}^{g_0A_{\mu}(\bm{x}+\frac{\hat{\mu}}{2})}$ of the link variables,
the action is expanded as
\begin{eqnarray*}
\mathcal{L}&=&\mathrm{Tr}(U_{\mathcal{C}}) + \overline{\psi}\Gamma U_{\mathcal{D}}\psi\\
&=&\sum_r \frac{g_0^r}{r!} V^{a_1\cdots a_r}_{\mu_1\cdots\mu_r}
         A^{a_1}_{\mu_1}\cdots A^{a_r}_{\mu_r} 
+\sum_r \frac{g_0^r}{r!} \overline{\psi}^{b} \tilde{V}^{a_1\cdots a_r,bc}_{\mu_1\cdots\mu_r}
         A^{a_1}_{\mu_1}\cdots A^{a_r}_{\mu_r} \psi^{c}
\end{eqnarray*}
with {vertex functions}
\begin{eqnarray*}
V^{a_1\cdots a_r}_{\mu_1\cdots\mu_r} (k_1,\ldots,k_r) &=& 
\mathrm{Tr}\left(t^{a_1}\cdots t^{a_r}\right) \times
\sum_{i} f_{i} \mathrm{e}^{i \sum_j k_j\cdot v_{i,j}}\\
\tilde{V}^{a_1\cdots a_r,bc}_{\mu_1\cdots\mu_r} (k_1,\ldots,k_r;p,q)&=& 
\left(t^{a_1}\cdots t^{a_r}\right)_{bc} \times
\sum_{i} f_{i} \Gamma_{\alpha_{i}} \mathrm{e}^{i \left(p\cdot x_i +q\cdot y_i +\sum_j k_j\cdot v_{i,j}\right)}
\end{eqnarray*}
giving the Feynman rules.

The algorithm for achieving this expansion starts from the encoding
of individual terms in the vertex function as so-called
{entities}
\[
E_i = \left(\bm{x}_i,\bm{y}_i;\bm{v}_{i,1},\ldots,\bm{v}_{i,r};\alpha_i\right)
\]
each of which carries an amplitude $f_i$.
The crucial property of entities is the multiplication rule
\begin{eqnarray*}
E E' &=& \left(\bm{x},\bm{y}'+\bm{c};\bm{v}_{1},\ldots,\bm{v}_{r},\bm{v}'_{1}+\bm{c},\ldots,\bm{v}'_{s}+\bm{c};\alpha_{k}\right)\\
\end{eqnarray*}
where $\bm{c}=\bm{x}'-\bm{y}$ and $\alpha_k$ is defined via
$\Gamma_{\alpha_i}\Gamma_{\alpha_j}=\phi_{\alpha_i\alpha_j}\Gamma_{\alpha_k}$.
Entities differing by only translations are {equivalent}
by momentum conservation.
Additional structure (e.g. a non-trivial colour structure) can also be encoded
by adding additional fields to the entity and amending the entity algebra
accordingly.

A field is then defined as a double mapping
\[
F=\left\{ \left({\mu_1,\ldots,\mu_r}\right) \mapsto \left\{ E_i \mapsto f_i \right\}\right\}
\]
which encodes a generic Wilson line, and multiplication of field objects
is defined accordingly in terms of entity multiplication by
\begin{eqnarray*}
F F' &=& \left\{ \left({\mu_1,\ldots,\mu_r,\mu'_1,\ldots,\mu'_s}\right) \mapsto \left\{ E E' \mapsto C_{rs} \phi f f' \right\}\right\}
\end{eqnarray*}
where $C_{rs}=(r+s)!/(r!s!)$ is a combinatorial factor and $\phi$ is the phase
from the multiplication of the spin matrices belonging to the entities
$E$ and $E'$, as defined above. Addition of field objects is defined by the
addition of the amplitudes belonging to the individual entities, with the
amplitude in the sum of an entity present in only one of the summands being
set to its amplitude in that summand.

The basic building block from which smeared links, operators and actions are constructed in an iterative fashion is the {simple link} encoded as
\begin{eqnarray*}
U_\mu(\bm{x}) &=& \mathrm{e}^{g_0A_\mu(\bm{x}+\frac{\hat{\mu}}{2})}
= \left\{ \left({\mu,\ldots,\mu}\right) \mapsto \left\{ \left(\bm{0},\hat{\mu};\frac{\hat{\mu}}{2},\ldots,\frac{\hat{\mu}}{2};0\right) \mapsto 1 \right\}\right\}\,,
\end{eqnarray*}
and predefined building blocks (e.g. smeared links, covariant derivatives and field strength tensors) constructed from this are provided as part of HiPPy.

\subsection{The HPsrc package}

The HPsrc package consists of a suite of Fortran~95 modules complementing HiPPy. While the output of HiPPy is in principle suitable to being converted directly into an analytic form, this is not usually necessary or even useful in lattice perturbation theory. We therefore have implemented routines that read in the HiPPy-generated Feynman rules at runtime and use them to construct the vertex functions and propagators for given momenta on the fly. This also offers the great advantage of being able to write Feynman diagrams in an action-blind way, so as to be able to recompute the same quantities for another action without needing to write new code. HPsrc also provides facilities for automated differentiation of Feynman diagrams, so that neither analytic manipulations nor inaccurate numerical derivatives are needed.

\section{Incorporating Background Fields}

The background field technique has long been known as a valuable tool in field theory. In
\cite{LW:BGF1},
L\"uscher and Weisz showed that the theorem about dimensional regularisation stating that renormalisation of the effective action does not require additional counterterms beyond those needed for the renormalisation of the action holds also in the case of lattice gauge theory. This makes it possible to use the background field technique to perform calculations such as relating the bare lattice coupling to the $\overline{\textrm{MS}}$ coupling
\cite{LW:BGF2}.
To determine the coefficient of the $\bm{\sigma\cdot B}$ term in the (moving) NRQCD action
\cite{wip},
only the background field technique can guarantee the gauge invariance of higher-dimensional operators which will necessarily be generated in an effective theory such as (m)NRQCD. This makes it desirable to incorporate support for the background field method into the HiPPy/HPsrc packages.

\subsection{Background fields in HiPPy}

We decompose our fields into a background field $B_\mu$ and quantum
fluctuations $q_\mu$ by parametrising the basic gauge link as
\[
U_{\mu}(\bm{x})=\mathrm{e}^{g_0 q_{\mu}(\bm{x}+\frac{\hat{\mu}}{2})} \mathrm{e}^{B_{\mu}(\bm{x}+\frac{\hat{\mu}}{2})}
\]
We note that this does not affect the combinations of $\bm{x}$, $\bm{y}$, $\bm{v}_{i,j}$ that are possible, and that hence the entity algebra remains unaffected. However, the gluon fields living at each lattice point $\bm{v}$ can now be either {quantum} or {background}, with the quantum fields always appearing to the left of the background fields coming from the same link, and thus the field objects need to keep track of nature of individual gluon fields.

This leads to a new mapping structure for field objects given now by
\[
F=\left\{ \left({\mu_1,\ldots,\mu_r}\right) \mapsto \left\{ E_i \mapsto (f^{q\cdots q}_i, \ldots, f^{B\cdots B}_i)\right\}\right\}
\]
with an order-$r$ entity being mapped to an $2^r$-tuple of amplitudes.

With this structure, multiplication of fields now assigns
\[
f^{x_1\cdots x_ry_1\cdots y_s}_k:=C_{rs} \phi f^{x_1\cdots x_r}_i f^{y_1\cdots y_s}_j
\]
where $C_{rs}$ and $\phi$ are defined as before, and the simple link becomes
\[
U_\mu(\bm{x}) = \mathrm{e}^{g_0A_\mu(\bm{x}+\frac{\hat{\mu}}{2})}
= \left\{ \left({\mu,\ldots,\mu}\right) \mapsto \left\{ \left(\bm{0},\hat{\mu};\frac{\hat{\mu}}{2},\ldots,\frac{\hat{\mu}}{2};0\right) \mapsto (1,\ldots,f^{q\cdots qB\cdots B},\ldots,1) \right\}\right\}\,,
\]
where
\[
f^{q^sB^{r-s}}=\frac{r!}{s!(r-s)!}
\]
and all other $f^x$ vanish identically.
This definition maintains the ordering of the background and quantum fields throughout the expansion procedure. We note that this precludes performing any symmetrisation over the gluon fields at this stage, and that in background gauge all symmetrisation is to be deferred to the evaluation of the Feynman diagrams.

\subsection{Background field gauge in HPsrc}

In order to support background field calculations, we also need to extend HPsrc so that it supports fixing to background field gauge.

The {gauge fixing} term for this is
\[
\mathcal{L}_{gf} = -\frac{1}{2\xi} \mathrm{Tr}\left(D_{\mu}^*q_{\mu}\right)^2
\]
with
\[
D_\mu^*f(x) = \left[f(x)-\rme^{-B_\mu(x-\frac{\hat{\mu}}{2})}f(x-\hat{\mu})\rme^{B_\mu(x-\frac{\hat{\mu}}{2})}\right] \,,
\]
which gives rise to additional terms in all purely gluonic vertices with exactly two quantum fields. These terms have been implemented in HPsrc for the propagator and three-gluon vertex.

Similarly, the Fadeev-Popov {ghost action} in background field gauge involves background covariant derivatives $D_{\mu}$ instead of finite differences, and $\mathrm{Ad}(q_{\mu})$ instead of $\mathrm{Ad}(A_{\mu})$.

\section{Accelerated Automatic Differentiation}

In HPsrc, we use {TaylUR}
\cite{TaylUR}
for the {automatic} computation of derivatives with respect to an external momentum $\bm{P}$. In this way, derivative information is automatically propagated from vertices and propagators to Feynman diagrams and quantities constructed therefrom, but in order to save computation time, the derivatives of the vertex functions themselves are constructed explicitly in the HPsrc code.

For a momentum decomposed as $\bm{k}_j=\bar{\bm{k}}_j+r_j\bm{P}$, the derivative of the reduced vertex reads
\[
\!\!\!\partial^{n_1}_{P_1}\cdots\partial^{n_D}_{P_D}Y =
\sum_i f_i {\prod_{\mu=1}^D \left(i \sum_j r_j v_{i,j,\mu}\right)^{n_\mu}} \!\!\!\mathrm{e}^{i \sum_j k_j\cdot v_{i,j}}
\]
We note that the $\prod\left(\ldots\right)$ term above is {momentum-independent}, and needs to be computed only {once} in order to compute the derivatives for each momentum in a set of momenta with identical routings $r_j$; only the exponential needs to be recomputed for each momentum. It is therefore possible to achieve a significant speed-up by {momentum vectorisation}, such that the vertex function is called with a vector of momenta with identical routings $r_j$ and returns a vector of values.

\section{GPU acceleration of Reduced Vertices}

The main work (about 85\%) in a perturbative calculation using HPsrc
comes from evaluating the reduced vertex. This function is
particularly suitable for parallelisation using a General Purpose GPU
(GPGPU). Data transfer times are small and the momentum vectorisation
discussed above makes the overhead per momentum point negligible (as
the monomial data can be reused).

The reduced vertex routine was extracted as a separate kernel for
testing. Derivatives of the reduced vertex were not initially
calculated. The kernel consists of a two-level loop nest. The outer
loop (over independent momentum points with loop index $n$) is
trivially parallel. The inner loop (the sum over monomials with loop
index $i$) contained some dependencies. We fixed the number of
monomials at 8000 (representative of the HISQ action) and varied the
number of points.

The kernel was accelerated on an NVIDIA Fermi C2050 GPU in two
ways. First, a corresponding CUDA kernel was written and called via a C
wrapper. Secondly, the original Fortran90 was accelerated
using the directives implemented in the PGI compiler. 

Initially, only the outer loop was parallelised, explicitly
in the CUDA and automatically by the PGI compiler. In both cases, GPU
performance was significantly increased by reordering the $k$ array so
that index $n$ was the fastest-varying in memory. This allowed
``coalescing'' of loads from the GPU global memory.

For a wide range of problem sizes, the CUDA kernel (including data
transfers) was around 52 times faster than the best version of the kernel
running on a single core of an Intel Nehalem processor. The PGI
directive version was only 20\% slower than the CUDA kernel.

\begin{figure}
\centering
\includegraphics[width=5in]{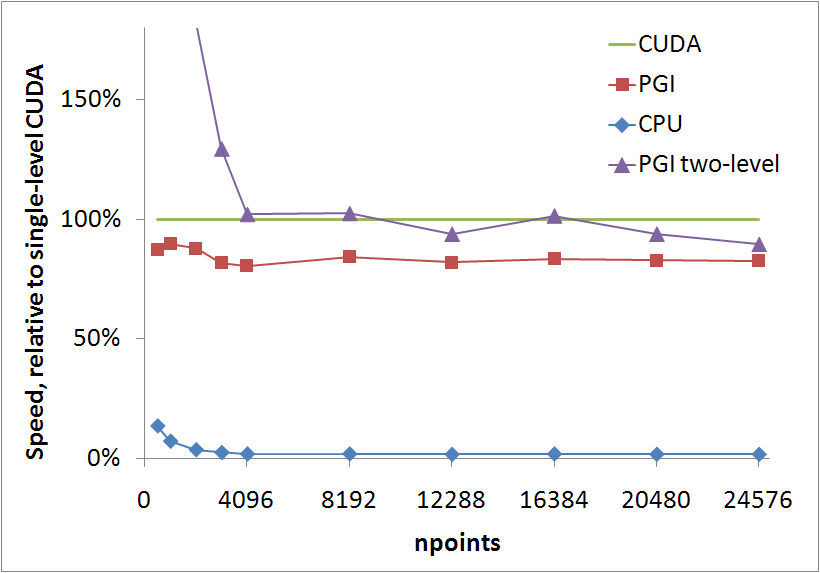}
\caption{\label{fig_hpsrc_gpu}Comparing the performance of the HPsrc
  kernel on an NVIDIA Fermi GPU for increasing numbers of momentum
  points.}
\end{figure}

Further performance gains were obtained by restructuring the kernel so
that loops over both $n$ and $i$ could be parallelised. When this was
done in the PGI directive-accelerated kernel, the performance matched
that of the original CUDA version. The same refactoring in the CUDA is
currently in progress. Our results are summarised in
Fig.~\ref{fig_hpsrc_gpu}.

This work is now being ported into the main application.

\section{Summary}

The {HiPPy/HPsrc} package has been extended to enable calculations in background field gauge. The new functionality will be used to calculate to $\mathcal{O}(\alpha_s)$ corrections to the coefficients of the (m)NRQCD action.

The re-use of common routing information enables a significant speed-up of calculations involving automatic differentiation of vertices. A further speed-up can be achieved by rewriting the generic vertex functions as {CUDA} kernels. An optimised implementation is in preparation.

\end{document}